\documentclass[conference]{IEEEtran}
\IEEEoverridecommandlockouts

\usepackage{cite}
\usepackage{amsmath, amssymb, amsfonts}
\usepackage{algorithmic}
\usepackage{graphicx}
\usepackage{textcomp}
\usepackage[normalem]{ulem}
\usepackage[usenames, dvipsnames, table]{xcolor}
\usepackage{pgfplots}
\pgfplotsset{compat=1.17}
\usepackage[inkscapelatex=false]{svg}
\usepackage{tabularx}
\usepackage{colortbl}

\graphicspath{{/app/}}

\usepackage{xspace}

\usepackage{array}
\newcolumntype{P}[1]{>{\centering\arraybackslash}p{#1}}

\def\BibTeX{{\rm B\kern-.05em{\sc i\kern-.025em b}\kern-.08em
    T\kern-.1667em\lower.7ex\hbox{E}\kern-.125emX}}

\usepackage{fancyhdr}
\fancypagestyle{firstpage}{%
 \fancyhf{}%
 \fancyhead[C]{2024 5th CPSSI International Symposium on Cyber-Physical Systems (Applications and Theory) (CPSAT)}
}

\begin{document}

\title{Uncovering EDK2 Firmware Flaws: Insights from Code Audit Tools\\}

\author{Mahsa Farahani\textsuperscript{*}, Ghazal Shenavar\textsuperscript{*}, Ali Hosseinghorban\textsuperscript{*}, Alireza Ejlali\textsuperscript{*}
\\
\textsuperscript{*}Embedded Systems Research Laboratory (ESRLab),\\Department of Computer Engineering, Sharif University of Technology, 
Tehran, Iran
\\
\{mahsa.farahani, ghazal.shenavar, ali.hosseinghorban1394, ejlali\}@sharif.edu
}

\maketitle
\thispagestyle{firstpage}  

\begin{abstract}

Firmware serves as a foundational software layer in modern computers, initiating as the first code executed on platform hardware, similar in function to a minimal operating system. Defined as a software interface between an operating system and platform firmware, the Unified Extensible Firmware Interface (UEFI) standardizes system initialization and management. A prominent open-source implementation of UEFI, the EFI Development Kit II (EDK2), plays a crucial role in shaping firmware architecture. Despite its widespread adoption, the architecture faces challenges such as limited system resources at early stages and a lack of standard security features. Furthermore, the scarcity of open-source tools specifically designed for firmware analysis emphasizes the need for adaptable, innovative solutions.

In this paper, we explore the application of general code audit tools to firmware, with a particular focus on EDK2. Although these tools were not originally designed for firmware analysis, they have proven effective in identifying critical areas for enhancement in firmware security. Our findings, derived from deploying key audit tools on EDK2, categorize these tools based on their methodologies and illustrate their capability to uncover unique firmware attributes, significantly contributing to the understanding and improvement of firmware security.
\end{abstract}


\section{Introduction}
Firmware acts as software coded directly into hardware across various devices, from personal computers to critical infrastructure components. It runs first when a device is powered on, initializing hardware and preparing for higher-level software operations \cite{9878283}. Initially introduced to reside in hardware's Read-Only Memory (ROM), firmware now plays a vital role in both personal computers and embedded systems as technology evolves \cite{10.1145/3132300.3132337}. UEFI is a widely adopted firmware standard, responsible for booting the operating system and configuring hardware, including the CPU and RAM. Its critical role in the boot process and widespread deployment across millions of devices makes UEFI a key component of modern computing systems \cite{6737940}.

Firmware security is often neglected despite its widespread use, leaving vulnerabilities overlooked. High-profile incidents like Mosaic Regressor \cite{mosaic}, Black Lotus \cite{blacklotus}, and LogoFAIL \cite{logofail} highlight the urgent need for stronger UEFI security. These threats can originate from any part of the supply chain, from silicon coding to secondary suppliers, and the long cycles needed to implement fixes mean that vulnerabilities remain exposed for extended periods. The need for robust firmware security is pressing, as vulnerabilities can have catastrophic consequences. Attacks exploiting these vulnerabilities could disrupt not just individual devices but entire systems or networks, potentially taking control of or disabling critical infrastructure components across a network, leading to widespread operational disruptions.
Furthermore, there are no security applications like antivirus and firewalls to protect the firmware, making it more vulnerable than the operating system \cite{NADIR2022100552}.

UEFI is just a specification, and the EFI Development Kit II (EDK II) is the most prominent open-source implementation of it \cite{edk2}.
The standardization provided by the UEFI specification and the EDK2 has expanded the scope of firmware development, allowing for the integration of more complex features traditionally associated with operating systems. These features are typically implemented through UEFI applications and drivers that extend the basic UEFI functionality. While this evolution has enhanced firmware flexibility and manageability, it also introduces greater security risks.

Despite the critical role of UEFI and EDK2 in modern computing, research in this area remains surprisingly underexplored, presenting a unique opportunity for valuable contributions. While some studies have examined specific UEFI vulnerabilities, none have focused on systematically applying static analysis tools to firmware, particularly the EDK2 platform. Key issues in the UEFI Image Loader have been identified, and a formally verified alternative has been proposed to improve security \cite{9394099}. Another work offers a comprehensive analysis of UEFI security challenges, classifying threats and offering recommendations \cite{6737940}. However, these studies lack a structured approach for applying auditing tools or a detailed taxonomy of these tools based on their methodologies.

Our research fills this gap by introducing a practical methodology for identifying vulnerabilities in EDK2 firmware, advancing beyond the theoretical models of previous works such as \cite{9345103}. While prior studies focus on specific vulnerability types \cite{9833723}, our approach categorizes code audit tools according to their methods, offering a broader assessment of security threats and actionable insights for mitigation.

This paper presents a novel taxonomy of code audit tools tailored to firmware, helping to select the appropriate tools for specific vulnerabilities or project types. To our knowledge, this is the first study to systematically classify and apply these tools to the unique challenges of EDK2 firmware security, marking a significant advancement in this underexplored field. Our contributions include:

\begin{itemize}

\item 
We introduce a novel classification system for code audit tools, detailing their unique methodologies and highlighting their applicability to firmware analysis.
\item 
Through the application of various code audit tools to the EDK2 project, we identified critical vulnerabilities such as buffer overflows and memory management flaws. Our analysis provides practical, actionable insights into how these vulnerabilities can be mitigated to enhance the security of EDK2 firmware.
\item 
We applied our findings to real-world scenarios within the EDK2 project, using specific examples of identified vulnerabilities to illustrate their security implications. These examples serve to validate our approach and demonstrate the practical impact of the identified issues on EDK2's overall security.
\end{itemize}

The remainder of this paper is organized as follows:  Section~\ref{sec:codeAudit} introduces a classification of code audit tools, emphasizing their distinct approaches and relevance to firmware analysis. Section~\ref{sec:experiments} details the findings from applying these tools to EDK2, including identified vulnerabilities and their security implications. The paper concludes in Section~\ref{sec:conclusion}.

\section{Code Audit}
\label{sec:codeAudit}
\newcommand{\mycomment}[1]{}

Code Audit is the detailed examination of source code within a programming project aimed at identifying bugs, potential security vulnerabilities, or any deviations from programming standards. This thorough evaluation is critical for uncovering issues that might lead to system failures, data breaches, or security incidents, such as the use of insecure functions or coding practices that make systems vulnerable to attacks \cite{7373782}. Despite the availability of safer alternatives, Unsafe function usage in major projects like Curl and Bitcoin \cite{Crifasi2018Cloud} underscores the need for proactive issue detection. Code Audits also delve into performance bottlenecks and maintenance challenges, ensuring that the codebase scales well, follows best practices, and maintains thorough documentation. This level of scrutiny is essential for the sustainable development of software, significantly enhancing the longevity and security of the application.

In the realm of code analysis, we differentiate between two primary types:
\begin{itemize}
\item \textbf{Static Analysis} Tools, which review code during its development before it is executed. This method is effective for assessing code quality and security but may flag theoretical issues that do not impact the system under actual operating conditions\cite{Fatima2018Comparative}.
\item \textbf{Dynamic Analysis} Tools, which assess code in its runtime environment, ensuring accurate detection of issues as they occur. Despite their accuracy, the applicability of dynamic tools in firmware environments is limited due to their high resource demands and the potential impact on system performance \cite{Gomes2009AnOO}.
\end{itemize}

\begin{figure}[ht]
    \centering
    \includegraphics[width=1.03\columnwidth]{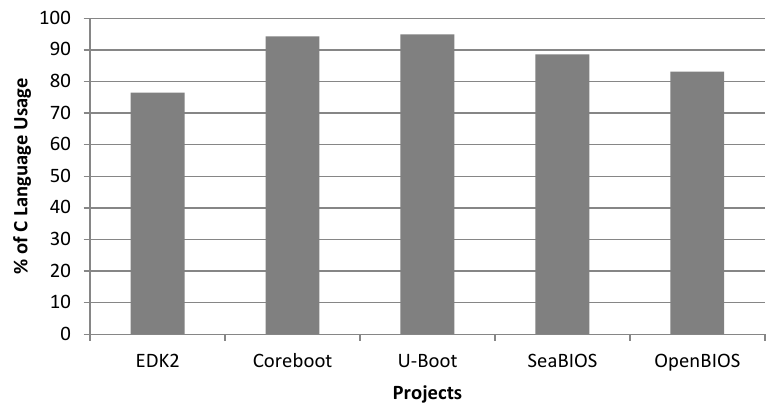}
    \caption{C Language Usage in Prominent Firmware Projects  \cite{edk2,coreboot,uboot,seabios,openbios}.}
    \label{fig:1}
\end{figure}

While there are numerous tools available for auditing various types of software code, this paper focuses on those suitable for firmware development. We investigate only static analysis approaches, as dynamic techniques are impractical in the resource-constrained environments typical of firmware development. It's important to note that firmware initializes the main memory; before this, it can only use the system’s registers and cache \cite{banik2022system}, leaving insufficient memory for dynamic auditing techniques to monitor code behavior and detect vulnerabilities. Furthermore, we concentrate on auditing tools that work with the "C" language. "C" offers extensive access to hardware, memory, stack, heap, etc., making it ideal for firmware development that requires the initialization of various system modules. However, this level of access also makes "C" programs particularly susceptible to security vulnerabilities, highlighting the importance of rigorous code auditing.
 Fig \ref{fig:1} illustrates that the majority of prominent firmware projects, as determined from our analysis of various open-source repositories, are developed using the "C" language.

In our analysis, we categorize code auditing tools into four distinct groups, each tailored to address specific aspects of firmware security: String Matching, Execution Flow, Symbolic Execution, and Machine Learning. 
\textbf{String Matching} tools are crucial for firmware auditing as they quickly identify known patterns of vulnerabilities in the code, allowing for rapid detection of common security issues typical in firmware.
\textbf{Execution Flow} analysis is particularly important for firmware because it reveals how data is processed and moved within these systems, crucial for detecting unusual sequences or operations that could lead to security breaches, especially in systems that may not have extensive logging like larger operating systems do.
\textbf{Symbolic Execution} proves highly beneficial for firmware due to its ability to model the constrained operational conditions of firmware. It abstracts operations into symbolic representations, enabling comprehensive analysis of all possible execution paths and edge cases, thus revealing hidden vulnerabilities that static methods might overlook in the specialized nature of firmware.
\textbf{Machine Learning} techniques are essential in firmware security for predicting and identifying both familiar and new threats by analyzing patterns from large datasets of past security incidents and firmware behavior. This approach is vital for adapting to the dynamic nature of threats in the firmware landscape, where traditional methods may not suffice.

These categories form the foundation of our methodology, significantly enhancing firmware protection against current and emerging threats. In the following subsections, we introduce these categories and highlight a key code audit tool that leverages this method.

\subsection{String Matching}
In our study, we identify string matching as an effective technique in code analysis, particularly valuable for detecting potential security vulnerabilities in firmware code. This method, which utilizes regular expressions, adeptly targets insecure or suspicious patterns, a key to preventing future security breaches in firmware systems. Through the use of specific signatures, string matching tools are equipped to thoroughly examine the codebase, uncovering instances of these critical patterns. Highlighting this application of string matching within firmware development, we underscore its significant role in enhancing the early detection of security risks and its pivotal contribution to improving firmware security management.

Furthermore, this technique enhances the efficiency of code audits by comparing code differences and maintaining updated lists of code structures, which are essential optimizations in firmware development where computational resources are limited \cite{banik2022system}. While false positives present a challenge with string matching tools, their strategic application in firmware code analysis can yield substantial benefits, especially when paired with complementary methodologies to reduce inaccuracies.

Through our research, we have incorporated Rough Auditing Tool for Security (RATS) \cite{RATS2009} and FlawFinder \cite{flawfinder2023}, leading open-source tools for code analysis, applying them to the EDK2 framework. The choice to rely primarily on string matching tools is further justified by the standardized and repetitive sequences often found in firmware code due to its close hardware integration. This predictability makes string matching particularly effective, allowing for rapid identification of known vulnerable patterns across different firmware versions and platforms. Moreover, the direct application of string matching aligns perfectly with the resource constraints and security needs of firmware systems, enabling quick vulnerability assessments without the overhead of more computationally intensive techniques. This focused approach ensures that our security analysis remains both thorough and practical, addressing the immediate needs of firmware security with proven, efficient tools.

\begin{itemize}
\item 
\textbf{RATS} is a specialized static analysis tool that enhances security by using string matching to uncover potential vulnerabilities \cite{RATS2009}. It evaluates code across commonly used languages such as Python, PHP, Perl, C/C++, and Ruby, ensuring efficiency and the ability to handle large codebases effectively. While RATS is versatile enough to be applicable in various programming environments, its robust detection capabilities make it useful for examining more specific applications such as firmware. It sorts identified vulnerabilities by severity, providing insights into risks like memory overflow attacks and Time of Check Time of Use (TOCTOU) race conditions. Detailed reports include explanations of vulnerabilities and precise locations within the code. RATS has been shown to demonstrate a reasonable success rate in detecting security vulnerabilities, effectively identifying critical errors such as buffer overflow, insufficient control of resource identifiers, and OS command injection \cite{DBLP:journals/corr/abs-1805-09040}.

\item 
\textbf{Flawfinder} stands out in the realm of static analysis by employing string matching to scrutinize C, C++, and header files for security vulnerabilities  \cite{flawfinder2023}. This Python-based tool, prioritizing thoroughness albeit at a slower pace compared to similar tools like RATS, leverages the Common Weakness Enumeration (CWE) framework. CWE's comprehensive categorization enhances Flawfinder's ability to provide detailed, actionable insights into identified risks, ranking them from high to low severity and associating each with CWE identifiers for standardized reference. A distinctive feature of Flawfinder is its integration with the Software Assurance Reference Dataset (SARD) \cite{article} and its subset, Juliet, which furnishes a broad spectrum of test cases for in-depth analysis. When operated at its highest severity level, Flawfinder demonstrates a strong focus on precision, effectively reducing false positives and ensuring that critical vulnerabilities are prioritized. In the SARD test cases, Flawfinder successfully identified key vulnerabilities, reinforcing its role as an essential tool for thorough security assessments in large codebases\cite{10.1007/978-3-030-83723-5_13}. The tool's output is meticulously detailed, offering statistics on the volume of code analyzed, the severity of detected errors, and specific CWE identifiers, alongside concise descriptions for each vulnerability.
\end{itemize}

\subsection{Execution Flow}
Execution Flow Analysis is proposed as a novel and comprehensive approach for code auditing, focusing on both control and data flows within applications to uncover potential security vulnerabilities. This method is particularly adept at detecting a range of issues, including unauthorized data accesses, improper variable type conversions, memory overflows, and the presence of unused variables. By meticulously tracing the execution path of a program, it assesses the use of insecure functions and verifies adherence to necessary security standards, ensuring safe operation.

In the specific context of firmware security, Execution Flow Analysis proves invaluable. Firmware's predictable yet intricate interactions with hardware demand an exhaustive understanding of control flows to identify and mitigate complex vulnerabilities that simpler analysis might miss. This method's ability to trace data and command paths through the system reveals both overt and subtle security flaws, which are crucial in preventing exploits that could severely impact system stability and security.

Our focus on Execution Flow Analysis is driven by its capacity to deliver a detailed view of firmware behavior under real operational conditions. It provides a robust framework for evaluating how firmware interacts with hardware, making it possible to pinpoint and rectify security weaknesses efficiently. This analytical method forms a critical part of our security assessment strategy, particularly in the examination of firmware where non-standard code structures and direct hardware interactions are common. Cppcheck is the tool we utilize to enhance our analysis, capitalizing on its ability to inspect C and C++ codebases effectively \cite{cppcheck2023}.

\begin{itemize}
\item
\textbf{CppCheck}, a static analysis tool designed for C and C++ code, plays a critical role in enhancing security across various development projects, including firmware. Crafted in C++, Cppcheck is available in both free and commercial versions to suit diverse development requirements. Its proficiency in analyzing non-standard code structures makes it especially valuable in contexts like firmware security, where unconventional and complex code patterns frequently occur. CppCheck performed well, achieving an F-Score of 0.78 in a comprehensive evaluation of Static Application Security Testing (SAST), demonstrating its effectiveness in detecting vulnerabilities in C code\cite{10.1007/978-3-030-83723-5_13}.
\end{itemize}

\subsection{Symbolic Execution}
Symbolic Execution represents a sophisticated approach in the domain of software testing  particularly beneficial in firmware development, where ensuring the utmost correctness and reliability is paramount\cite{10.1145/3182657}. This technique conceptualizes variables as symbols subject to defined constraints, enabling the execution of code across all conceivable symbol values. Such comprehensive exploration is instrumental in identifying errors and validating that the program is devoid of unexpected behavior, essentially providing a guarantee on the firmware's correctness.

Despite its unparalleled ability to assert program correctness, Symbolic Execution is marked by significant computational demands. It is inherently time-consuming and memory-intensive, with the complexity and resource requirements escalating exponentially alongside the increase in variables and states. This aspect poses a notable challenge, particularly in extensive firmware systems characterized by a multitude of variables and intricate control flows.

As a strategic response to the intensive demands of Symbolic Execution, Concolic Execution emerges as a viable alternative \cite{10.1145/2810103.2813663}. This method combines concrete value execution with symbolic analysis, directly executing the code while concurrently exploring unexecuted segments by satisfying the necessary conditions. Although Concolic Execution strives to encompass all possible code execution states, it is important to acknowledge that it might not capture certain edge cases or exceptional conditions, potentially leaving some vulnerabilities undetected.

Utilizing Symbolic and Concolic Execution in firmware analysis necessitates a significant allocation of resources, primarily memory, and time. The exhaustive nature of these analyses, while ensuring a high degree of correctness and coverage, demands careful consideration in their application, particularly in resource-constrained environments or within tight development timelines. Given the high resource demands and the potential for exponential complexity growth in firmware systems, our study has opted not to use Symbolic Execution. Our focus is on methodologies that balance depth of analysis with practical feasibility, particularly those that can be integrated more seamlessly into regular development cycles without the extensive overhead. This decision aligns with our goal to maintain a pragmatic approach in enhancing firmware security, ensuring that our methods are both effective and applicable within the typical constraints faced by firmware developers.

\subsection{Machine Learning}

Machine learning, with its extensive application range, has carved out a niche in the realm of code analysis  proving particularly beneficial for firmware development\cite{7855962}. A pivotal area within machine learning that has garnered attention for its utility in code analysis is Natural Language Processing (NLP). Traditionally aimed at deciphering human languages, NLP methodologies have been adapted for analyzing code, offering a unique lens through which firmware code structure can be understood. This adaptation leverages the compiler's front-end output to facilitate a deeper comprehension of firmware code structure, as detailed in related studies. These NLP-based tools undergo training and evaluation using vast repositories of code, enabling them to recognize patterns and recurring problems within new firmware codebases, which is invaluable for identifying common issues across different firmware projects\cite{9293321}.

Despite the potential benefits, machine learning and NLP in firmware code analysis have inherent limitations. The primary challenge is their inclination towards identifying previously encountered problems, limiting their ability to uncover new, undocumented vulnerabilities. This tendency, coupled with the significant resource requirements for training these tools and the complexity of maintaining the necessary infrastructure, restricts their utility for our research aims. Machine learning tools primarily excel at recognizing known issues within new firmware projects but struggle with novel vulnerabilities that are critical to address as firmware evolves in complexity.

Given these limitations and the need for methods that provide immediate, actionable insights, we opted not to use machine learning and NLP in our study. Our focus has been on more direct analytical techniques that align with the practical constraints of firmware development timelines and environments. This approach ensures that our security analysis remains effective, applicable, and responsive to the immediate needs of firmware developers without the extensive setup and learning curve associated with machine learning.

\begin{table*}[ht]
\centering
\caption{Comparison of Code Auditing Methods (Example tools:, \textit{KLEE} \cite{klee-github}, \textit{FIE} \cite{182944} \textit{CodeQL} \cite{codeql}, \textit{FUNDED} \cite{Wang_FUNDED_NISL})}
\resizebox{\textwidth}{!}{%
\begin{tabular}{|c|c|c|c|c|c|c|}
\hline
\rowcolor{lightgray} \textbf{Method} & \textbf{Memory Usage} & \textbf{Speed} & \textbf{Accuracy} & \textbf{Example Tools} & \textbf{Pros} & \textbf{Cons} \\ \hline
\textbf{String Matching} & Low & Fast & Moderate & RATS, FlawFinder & Fast, low resource & High false positives, missing complex cases \\ \hline
\textbf{Execution Flow} & Moderate & Variable & High & CppCheck & Detailed, finding logic errors & False positives, slower on large codebases \\ \hline
\textbf{Symbolic Execution} & Very High & Slow & Very High & KLEE, FIE & Comprehensive, deep detection & High resource use, complex setup \\ \hline
\textbf{Machine Learning} & Moderate & Fast & High & CodeQL, FUNDED & Adaptive, handling large data & High setup cost, limited by training data \\ \hline
\end{tabular}%
}
\label{comparison}
\end{table*}
\subsection{Hybrid Methods}
Within the realm of firmware analysis, leveraging a multifaceted approach can offer comprehensive insights by mitigating the drawbacks inherent to singular methods. Hybrid methods embody this strategy by amalgamating various techniques to enhance the depth and accuracy of code analysis.One such tool that utilizes this hybrid approach is Sys \cite{sys-github}. However, the application of these hybrid methods in our study was constrained by the unique challenges of firmware analysis. Specifically, firmware environments often feature highly specialized and proprietary systems that require customized tools for effective analysis. Hybrid methods, while beneficial in theory, require extensive adaptation and calibration to interface effectively with such specific systems. This adaptation process can significantly extend development time and complicate the scalability and reproducibility of findings. Therefore, our focus remained on more direct and established methods that ensured consistent applicability and reliable results within the specific constraints and complexities of firmware systems.

In concluding our exploration of various code analysis tools, our approach has been to strategically select and adapt tools not originally designed for firmware development, demonstrating their applicability and effectiveness within this specialized field. By integrating tools  alongside advanced methods like symbolic execution and machine learning-based analysis, we have tailored a comprehensive toolkit that addresses the unique challenges and intricacies of firmware security. A comparative summary of these methods, including their memory usage, speed, accuracy, and key strengths and weaknesses, is provided in Table \ref{comparison}. This curated selection of tools, augmented by our innovative application and methodology, significantly contributes to the advancement of firmware code auditing. It underscores our commitment to pushing the boundaries of conventional software analysis to meet the specific needs of firmware development, thereby enhancing the security and robustness of firmware systems.

\section{Experiments \& Results} 
\label{sec:experiments}

This section details the application and findings of three static analysis tools like RATS, Flawfinder, and CppCheck on the EDK2 project. Each tool identified various vulnerabilities, providing insights into the project's security posture and guiding the prioritization of mitigation efforts.

\subsection{RATS}
\begin{figure}[ht]
    \centering
    \includegraphics[width=1.03\columnwidth]{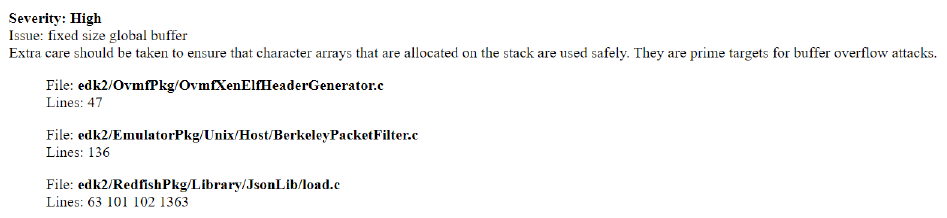}
    \caption{RATS output showing high-severity buffer overflow risks.}
    \label{fig:RATS}
\end{figure}

An application of RATS to the EDK2 project revealed 57 vulnerabilities classified into 14 categories, with a distinction between nine severe and five moderate categories. This segmentation aids in prioritizing security efforts, focusing on areas of highest risk first. As illustrated in Fig \ref{fig:RATS}, RATS provides a detailed breakdown of these vulnerabilities, highlighting the most critical issues that need immediate attention. Table \ref{rats} shows a summary of these vulnerabilities, categorized by their severity.

The vulnerabilities highlighted by RATS underscore the critical necessity of protection mechanisms in firmware environments. These issues, particularly buffer overflows and insecure API calls, confirm the effectiveness of these security measures in mitigating exploits that could compromise system integrity. RATS findings directly illustrate the risks associated with unchecked buffer sizes and improper API usage, reinforcing the importance of protection mechanisms. Blocking code execution in non-executable memory regions and preventing data corruption through buffer overflows demonstrates the value of these mechanisms in enhancing firmware security against sophisticated attacks.

Among the vulnerabilities identified, the issue of a fixed-size global buffer, as shown in Table \ref{rats}, is particularly concerning due to the absence of adequate bounds checking, which increases the risk of a buffer overflow. If exploited, this flaw could allow an attacker to send input data that exceeds the buffer's capacity, leading to an overflow that overwrites adjacent memory regions.
Such an exploit could have severe consequences, including the potential to alter the firmware's execution flow, allowing the attacker to execute arbitrary code during the boot process. This could result in the installation of persistent malware or the disabling of crucial security features, thereby compromising the system's security from the very start of its operation.

\begin{table*}[htbp]
\centering
\caption{Summary of Identified Vulnerabilities by RATS}
\begin{tabular}{|p{0.35\columnwidth}|p{1.4\columnwidth}|p{0.15\columnwidth}|}
\hline
\rowcolor{lightgray} 
\textbf{Vulnerability} & \textbf{Description} & \textbf{Severity} \\ \hline
Fixed size global buffer & Array stack usage, overflow risk. & Severe \\ \hline
strcpy and strcat & Second input validation, overflow prevention. & Severe \\ \hline
LoadLibraryEx & Full path specification, DLL injection avoidance. & Severe \\ \hline
EnterCriticalSection & Risky in memory constraints; InitializeCriticalSectionAndSpinCount preferred. & Severe \\ \hline
Fprintf & Safe resource inclusion, execution stability. & Severe \\ \hline
system/eval/compile & Dangerous element checking in first input. & Severe \\ \hline
Stat variable & Multiple lines after check, potential TOCTOU vulnerability. & Moderate \\ \hline
read and fgetc functions & Temporary memory boundary check recommended. & Moderate \\ \hline
realloc function & Memory not zeroed before use, unsuitable for security purposes. & Moderate \\ \hline
\end{tabular}
\label{rats}
\end{table*}

To mitigate this vulnerability, we recommend implementing robust bounds checking to ensure all input data is properly validated before processing. Additionally, adopting dynamic buffer allocation where input sizes are variable could further enhance security by reducing the likelihood of overflow conditions.
By addressing this vulnerability, the security of the EDK2 firmware can be significantly strengthened, reducing the risk of exploitative attacks during the critical boot process. This case was selected for its high potential impact, emphasizing the importance of tools like RATS in uncovering and addressing serious security vulnerabilities in firmware development.

While RATS does not ensure the detection of all errors, its ability to identify a wide range of potential vulnerabilities makes it a valuable tool in the firmware security auditing process. The specific findings from RATS analysis provide actionable insights, guiding the development of more secure firmware by addressing both common and complex security issues.

\subsection{Flawfinder}
In its application to the EDK2 project, Flawfinder identified a significant number of vulnerabilities, 1264 to be precise, underscoring the challenge of addressing security issues in large software projects. Given the volume of findings, the analysis strategically zeroes in on the use of insecure functions, providing a targeted approach to mitigating potential risks. Among these, the "strlen" function was highlighted as particularly prevalent, indicating a common area for improvement in secure coding practices. Flawfinder flagged this function because, if a string is not properly null-terminated, strlen could read beyond the intended buffer, leading to a buffer overflow. Such a scenario could be exploited by an attacker to overwrite critical memory, potentially injecting malicious code or disrupting the firmware's control flow, compromising the system during critical operations like booting.

To mitigate this risk, it is recommended to replace strlen with safer alternatives like strnlen, which includes bounds checking, and to enforce rigorous input validation to ensure all strings are properly null-terminated. By concentrating on specific vulnerabilities and their implications, Flawfinder aids in pinpointing critical security concerns, guiding developers towards safer programming methods. Its detailed reporting, grounded in the CWE framework, makes it an invaluable tool for enhancing the security posture of firmware projects like EDK2, despite the daunting task of navigating through a multitude of identified issues.

The vulnerabilities identified by Flawfinder underscore the necessity for adopting enhanced input validation routines, utilizing safer library functions, and implementing secure coding practices to mitigate identified risks. Specifically, addressing buffer overflow risks, ensuring secure file handling, and enhancing randomness in security functions are critical steps toward fortifying the EDK2 project against potential security threats. 
As detailed in Table \ref{function_usage}, the function usage counts, along with their associated CWE categories, highlight the urgent need for improvements. This table underscores the functions that contribute to specific security risks, marking areas requiring immediate attention.
The Flawfinder analysis, by shedding light on the various vulnerabilities within the EDK2 project, serves as a foundation for developing targeted interventions aimed at improving the project's overall security framework. Implementing the recommended mitigation strategies will be pivotal in addressing the identified vulnerabilities, thereby enhancing the security and reliability of the EDK2 project.

Given that RATS and Flawfinder use similar static analysis methods, we compared them to better understand their differences. RATS is more focused on identifying and prioritizing critical vulnerabilities with fewer false positives, while Flawfinder provides broader coverage of potential issues, though it may generate more false positives. This suggests that RATS is better for targeting severe vulnerabilities, whereas Flawfinder is more suited for comprehensive, detailed code reviews.

\begin{table}[htp]
\centering
\caption{Unsafe Function Counts and CWE Associations Identified by Flawfinder}
\begin{tabular}{|m{0.12\linewidth}|m{0.04\linewidth}|l||m{0.29\linewidth}|m{0.04\linewidth}|l|}
\hline
\rowcolor{lightgray} 
\textbf{Function} & \textbf{Cnt} & \textbf{CWE} & \textbf{Function} & \textbf{Cnt} & \textbf{CWE} \\
\hline
strlen            & 334           & 126          & atoi              & 5             & 190          \\
strcpy            & 161           & 120          & getenv            & 4             & 20, 807      \\
memcpy            & 150           & 120          & getc              & 4             & 20, 120      \\
char              & 144           & 119, 120     & fgetc             & 3             & 20, 120      \\
sprintf           & 100           & 134          & InitializeCriticalSection & 3        & --           \\
strcat            & 100           & 120          & printf            & 3             & 134          \\
fopen             & 95            & 362          & LoadLibraryEx     & 2             & 829, 20      \\
fprintf           & 85            & 134          & mkstemp           & 2             & --           \\
strncpy           & 23            & 120          & tmpnam            & 2             & 377          \\
strncat           & 15            & 120          & ShellExecute      & 2             & 78           \\
sscanf            & 10            & 20, 120      & srand             & 1             & 327          \\
vfprintf          & 7             & 134          & access            & 1             & 362          \\
snprintf          & 6             & 134          & vsprintf          & 1             & 134          \\
vsnprintf         & 5             & 134          & \_wtoi            & 1             & 190          \\
Read              & 5             & 20, 120      & chmod             & 1             & 362          \\
open              & 6             & 362          &                  &               &              \\
\hline
\end{tabular}
\label{function_usage}
\end{table}

\subsection{CPPcheck}
In our utilization of CppCheck, particularly the non-commercial variant, we analyzed the firmware codebase, which includes about 900,000 lines of C/C++ code, as measured using the tokei tool \cite{tokei}. Impressively, the analysis was completed in less than ten minutes. This swift performance is notable considering that, according to a referenced paper, CppCheck took hours to analyze 49 million lines, indicating that the relationship between the volume of code and the time required for analysis is not linear.
CppCheck classifies issues into six severity levels from 'Error' to 'Information,' allowing firmware developers to prioritize remediation efforts effectively. The categories include 'Error' for definite problems that require urgent attention; 'Warning' for potential runtime issues or undefined behaviors; 'Style' for highlighting problematic code structures to improve maintenance; 'Performance' for suggestions on optimizing code execution speed; 'Portability' for addressing compatibility issues across different platforms; and 'Information' for providing insights into tool usage and potential areas for improvement.

During the analysis of the EDK2 project, CppCheck identified 198 vulnerabilities, offering critical insights into the firmware's security by highlighting various severity-based vulnerabilities, excluding less critical structure and execution issues. It's noteworthy that CppCheck analyzes files individually and manages folders and file outputs in a way that could lead to false positives due to missing cross-file variable tracking. Here's a summary of the findings:

\textbf{Errors:} The bulk of the vulnerabilities, with 138 instances identified as errors, represent the most critical issues that require prompt attention.  These errors are highlighted in Table  \ref{cppcheck} as the most significant.
One particularly critical vulnerability identified by CppCheck involved the use of uninitialized variables within a memory management routine. This issue was flagged due to the inherent risks associated with uninitialized memory, where the variables may contain residual data from previous operations. Such behavior can lead to non-deterministic outcomes, including memory corruption, system instability, or even unintended control flow alterations. In a real-world scenario, an attacker could potentially exploit this vulnerability by leveraging the unpredictable state of uninitialized variables to trigger conditions that divert the firmware’s execution path, allowing for the injection of malicious code or the bypassing of security checks during the system startup process. Such an exploit could result in a full system compromise, undermining the integrity of the boot process and the overall security posture of the device.

To mitigate this vulnerability, it is imperative to ensure that all variables are explicitly initialized before use, thereby eliminating the unpredictability associated with uninitialized memory. Furthermore, adopting rigorous static analysis and dynamic testing practices can help detect and address instances of uninitialized variables early in the development lifecycle. Code reviews should also focus on enforcing initialization practices as part of secure coding standards. These mitigation strategies not only address the specific risks associated with uninitialized variables but also contribute to a more robust and secure firmware environment, reducing the likelihood of similar vulnerabilities being exploited in the future.

\begin{table}[htbp]
\centering
\caption{Summary of Identified Errors by CppCheck}
\label{cppcheck}
\begin{tabular}{|m{0.42\linewidth}|m{0.48\linewidth}|}
\hline
\rowcolor{lightgray} 
\textbf{Error Type} & \textbf{Description} \\
\hline
Memory leaks & Unreleased memory, risk to stability. \\
\hline
Invalid input to \texttt{putc()} & Out-of-range input values. \\
\hline
Uninitialized variables & Unpredictable behavior risk. \\
\hline
Syntax errors in ``if'' statements & Missing or wrong inputs. \\
\hline
realloc mistake & Unreleased memory, potential error. \\
\hline
Functions missing \texttt{va\_end} & Missing \texttt{va\_end}, undefined macros. \\
\hline
Pointer arithmetic issues & NULL dereference. \\
\hline
Uninitialized structure members & Incorrect or unintended behavior. \\
\hline
\end{tabular}
\end{table}

The analysis also flagged limitations in CppCheck's ability to assess relationships between different files, which may account for some of the issues related to uninitialized variables and pointer dereferencing.

\textbf{Warnings:} A smaller subset, comprising ten vulnerabilities, were classified as warnings. All ten warnings reported by CppCheck on EDK2 are of the type "Possible null pointer dereference." Zero address is considered an invalid address in most programs. However, in x86 systems, the zero address is valid in legacy BIOS because the 16-bit interrupt vector table (IVT) is at address zero. In current UEFI firmware, the zero address is always mapped. EDK2 includes a mechanism for NULL pointer protection. When this protection is enabled, the page table is updated to mark the address zero page as not present. This causes a Page Fault exception if any program tries to access the zero address, effectively catching null pointer dereferences.

\textbf{Notes:} Additionally, 49 vulnerabilities were categorized as notes. These suggest a lower likelihood of actual errors, often relating to the handling of null inputs in functions and variables, including arithmetic operations on NULL pointers.

\textbf{Information:} A singular finding was classified under information, highlighting a limitation of CppCheck in its analysis of more than 12 \texttt{ifdef} statements in a single file. This underscores the importance of using the ``--force'' option for a more thorough examination.

The vulnerabilities identified by CppCheck within the EDK2 provide critical insights into areas where security can be significantly improved. The high number of errors related to memory management, such as memory leaks and uninitialized variables, highlight the necessity for robust security mechanisms. Specifically, these issues underscore the importance of protection mechanisms, which can prevent the execution of potentially harmful code arising from such vulnerabilities. The detection of invalid inputs and pointer arithmetic issues further supports the need for Guard Pages, which help prevent memory overflows and ensure that such errors are caught early. The warnings related to possible null pointer dereferences illustrate the relevance of mechanisms that safeguard memory integrity, confirming the effectiveness of current strategies like mapped zero addresses in UEFI firmware.

\section{Conclusion}
\label{sec:conclusion}
This study has thoroughly analyzed firmware security within the EDK2, demonstrating the effectiveness of general code audit tools in identifying vulnerabilities. By applying these tools to EDK2, we have classified critical security flaws and provided insights into enhancing firmware protection.

Our novel classification of code audit tools offers a strategic framework for selecting appropriate security assessments, aligning tool capabilities with specific firmware vulnerabilities. This approach has validated the adaptability and effectiveness of these tools in a firmware context, suggesting a path forward for enhanced security measures.
The findings highlight the necessity of an ongoing, adaptive security assessment framework for firmware, especially as UEFI's complexity and functionality expand. Future research should focus on refining these tools for firmware-specific applications while developing automated solutions to anticipate and mitigate emerging threats.

In conclusion, our work advances the discussion on firmware security, providing practical solutions and theoretical insights that improve system resilience at a foundational level.

\bibliographystyle{IEEEtran}
\bibliography{refs}

\end{document}